\documentclass[epj]{webofc}
\usepackage[varg]{txfonts}   
\wocname{EPJ Web of Conferences}
\woctitle{CONF12}
\usepackage{float} 
\usepackage[caption=false]{subfig} 

\newcommand{\eqs}[1]{\begin{equation} \begin{split} #1\end{split} \end{equation} }
\renewcommand{\P}{{\cal P}}

\newcommand{\ie}{{\it i.e.}}

\usepackage{xspace}
\usepackage{comment}

\newcommand{\cf}[1]{{Fig.~\ref{#1}}}

\usepackage{hyperref}
\hypersetup{
  pdftitle={},%
  pdfauthor={},%
  pdfsubject={},%
  pdfkeywords={},%
  pdfstartview={},%
  bookmarksopen=true, breaklinks=true, debug=true, %
  colorlinks=true, linkcolor=blue, citecolor=blue, urlcolor=blue
}

\begin{document}
\selectlanguage{english}
\title{$J/\psi+Z$ production at the LHC}

\author{Jean-Philippe Lansberg\inst{1}\fnsep\thanks{\email{Jean-Philippe.Lansberg@in2p3.fr}} \and
        Hua-Sheng Shao\inst{2} 
}

\institute{IPNO, Universit\'e Paris-Saclay, Univ. Paris-Sud, CNRS/IN2P3, F-91406, Orsay, France
\and
           Theoretical Physics Department, CERN, CH-1211 Geneva 23, Switzerland 
}

\abstract{%
We briefly review recent results which we have obtained in the study
of $J/\psi+Z$ production at the LHC.  Considering our NLO computation
in the Colour Evaporation Model (CEM) as an upper theory limit
for the single-parton-scattering contributions, we claim 
that the existing data set from ATLAS points at a dominant
double-parton-scattering contribution with an effective 
cross section smaller than that for jet-related observables.
As a side product of our analysis, we have computed, for the first
time, the one-loop QCD corrections to the $J/\psi$ $P_T$-differential 
cross section in the CEM.
}
\maketitle

\section{Introduction}

Thanks to the high luminosities collected at the LHC and the Tevatron,
observing associated production of a quarkonium with a vector boson or a heavy quark 
is not any more uncommon. The same applies to quarkonium-pair production.
Indeed, nearly a dozen of experimental analyses~\cite{Aaij:2011yc,Aaij:2012dz,Aad:2014kba,Khachatryan:2014iia,Abazov:2014qba,Aad:2014rua,Aad:2015sda,Aaij:2015wpa,Abazov:2015fbl,ATLAS-CONF-2016-047,Khachatryan:2016ydm} are now available along with many relevant theoretical works. Some gave predictions before these analysis~\cite{Artoisenet:2007xi,Li:2008ym,Lansberg:2009db,Mao:2011kf,Kom:2011bd,Gang:2012ww,Gang:2012js,Gong:2012ah,Lansberg:2013qka,Lansberg:2013wva,Dunnen:2014eta,Lansberg:2015lva,Likhoded:2016zmk,Koshkarev:2016ket}; some helped at the interpretation of these results~\cite{Sun:2014gca,Lansberg:2014swa,He:2015qya,Baranov:2015cle,Borschensky:2016nkv,Shao:2016wor,Lansberg:2016rcx,Lansberg:2016chs,Lansberg:2016muq}. 
Let us emphasise that many of these theoretical works relied on automated tools adapted to quarkonium production. Let us cite {\sc Madonia}~\cite{Artoisenet:2007qm}, {\sc Helac-Onia}~\cite{Shao:2012iz,Shao:2015vga} and FDC~\cite{Wang:2004du}.
We focus here on the associated production of a $Z$ boson with a prompt\footnote{We just performed a similar
analysis~\cite{Lansberg:2016muq} of the non-prompt sample of ATLAS.} $J/\psi$ at the LHC.

\section{Colour Evaporation Model up to one loop in $\alpha_s$}

As announced, CEM predictions can be considered as a realistic upper theory value 
for a class of associated-production observables where the gluon fragmentation is expected
to be dominant. This is the case of $J/\psi+Z$, but obviously also of single-$J/\psi$ production
at large $P_T$.
In order to consistently fix  the required parameter for such an upper theory value for our $J/\psi+Z$ NLO analysis, 
we have performed in~\cite{Lansberg:2016rcx} the first one-loop analysis 
of the differential cross section of single-$J/\psi$ hadroproduction.

Let us first recall that the CEM can be seen as the application to quarkonium production~\cite{Fritzsch:1977ay,Halzen:1977rs} of 
the principle of quark-hadron duality. Quarkonium-production cross sections 
are obtained by integrating the cross section for $Q \bar Q$ pair production 
in an invariant-mass region where its hadronisation into a 
quarkonium is likely. This means that, in practice, one considers the range 
between $2m_Q$  and the threshold to produce open-heavy-flavour hadrons, referred to as $2m_{H}$. 
One then multiplies this partial heavy-quark cross section by 
a phenomenological factor which accounts for the probability, $\P_{\cal Q}$, that the
pair eventually hadronises into a given quarkonium state. 
Our computation then reduces to that of
\eqs{\sigma^{\rm (N)LO,\ direct/prompt}_{\cal Q}= \P^{\rm direct/prompt}_{\cal Q}\int_{2m_Q}^{2m_H} 
\frac{d\sigma_{Q\bar Q}^{\rm (N)LO}}{d m_{Q\bar Q}}d m_{Q\bar Q}.
\label{eq:sigma_CEM}}
Owing to the simplicity of the model, the direct or prompt yields are obtained 
from the same computation but with a different overall factor. Different
attempts to "improve" the model focusing on specific mechanisms~\cite{Edin:1997zb,Damet:2001gu,BrennerMariotto:2001sv,Ma:2016exq} 
have been made (see~\cite{Lansberg:2006dh} for a brief overview of some of them) but they do not seem 
to be the object of a consensus as far as their relevance is concerned.

\begin{figure}[hbt!]
\begin{center}
\subfloat[LO]{\includegraphics[width=7.cm]{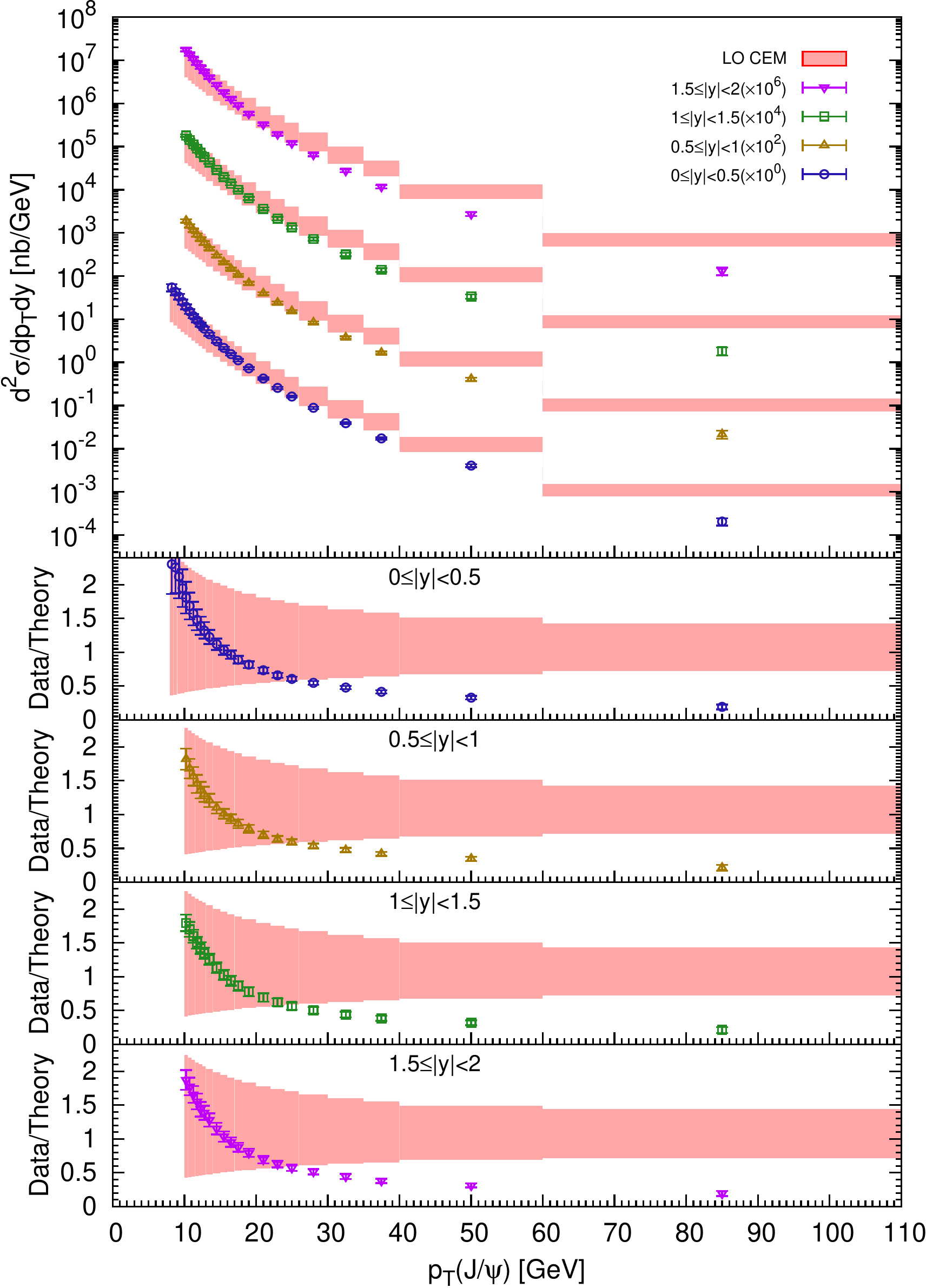}}
\subfloat[NLO]{\includegraphics[width=7.cm]{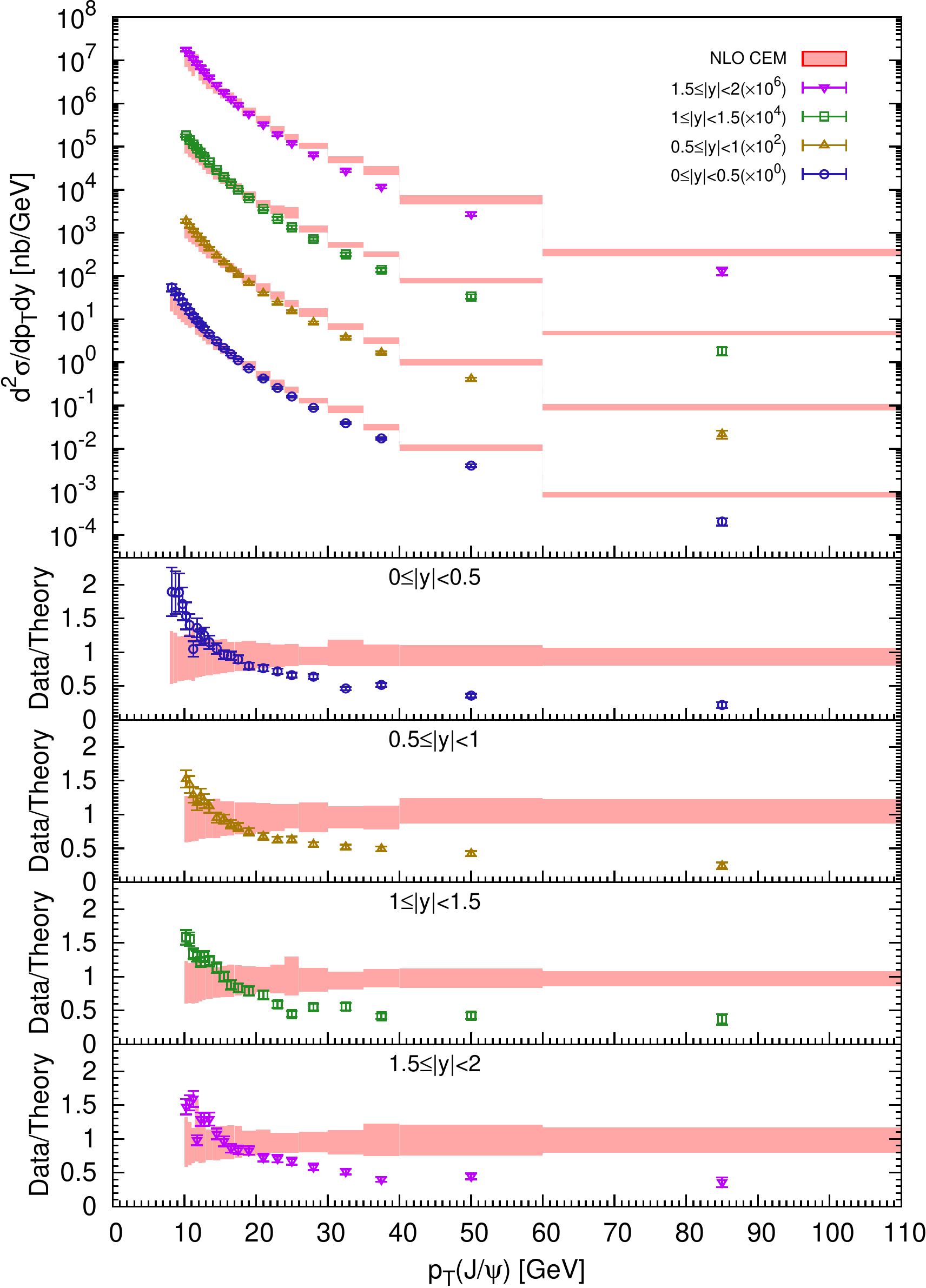}}
\caption{The ATLAS data~\cite{Aad:2015duc} compared to the CEM results for $d\sigma/dy/dP_T$ of 
$J/\psi$ + a recoiling parton at (a) LO and (b) NLO at $\sqrt{s}=8$ TeV. [The 
theoretical uncertainty band is from the scale variation (see the text)]. [Plots from~\cite{Lansberg:2016rcx}.] 
\label{fig:single-Jpsi-pT-spetrum}}
\end{center}
\end{figure}

A NLO comparison with $P_T$-integrated data from fixed-target experiments as well as from colliders was
performed in~\cite{Feng:2015cba}. The agreement was found to be satisfactory.
An interesting study of the relation between heavy-flavour and quarkonium 
production in the CEM can also be found in~\cite{Nelson:2012bc}. Along the lines of 
this analysis, we have decided to stick to $m_c=1.27$~GeV. In any case, there is a significant
correlation between the fit value of the phenomenological parameter $\P_{\cal Q}$ and $m_Q$; as such, 
the choice of the quark mass is probably less critical than for other processes
like open heavy-flavour production or quarkonium production in the colour-singlet model\footnote{See~\cite{Brodsky:2009cf,Lansberg:2010vq} for some examples.} for instance.

As what regards the $P_T$ spectrum of single $J/\psi$'s, 
the CEM is known to provide too hard a $P_T$ spectrum. However, before our study, 
such a statement was relying on studies using hard-scattering matrix elements at $\alpha_s^3$.
These are indeed NLO (one loop) for the  $P_T$-integrated yield but
not for the  $P_T$-differential cross section whose Born-order contribution 
is  at $\alpha_s^3$. It was therefore legitimate to wonder whether the 
 $P_T$ spectrum computed up to $\alpha_s^4$ would be different. 

Given the direct connection between the CEM and heavy-quark production, 
such a computation is in fact possible with modern tools of automated 
NLO frameworks, with some slight tunings. We 
have thus used {\small \sc MadGraph5\_aMC@NLO}~\cite{Alwall:2014hca} to perform our 
(N)LO CEM calculations for $J/\psi $ + a recoiling parton with a finite $P_T$. 
Since the heavy-quark mass dependence is de facto absorbed in the CEM parameter, the main theoretical 
uncertainties are coming from the renormalisation $\mu_R$ and 
factorisation $\mu_F$ scale variations which account for the unknown higher-order 
corrections. In practice,  we have varied them independently within 
$\frac{1}{2}\mu_0\le \mu_R,\mu_F \le 2\mu_0$ where the central scale $\mu_0$ 
is the transverse mass of the $J/\psi$ in $J/\psi$ + parton. We note the reduced theoretical
uncertainties at NLO.
We have used the NLO NNPDF 2.3 PDF set~\cite{Ball:2012cx} with $\alpha_s(M_Z)=0.118$ provided by 
LHAPDF~\cite{Buckley:2014ana}.

Fitting recent 8 TeV ATLAS data~\cite{Aad:2015duc} with $m_c=1.27$ GeV, 
we have obtained $\P^{\rm LO, prompt}_{J/\psi}=0.014\pm 0.001$ and
 $\P^{\rm NLO, prompt}_{J/\psi} =0.009 \pm 0.0004$. The $K$ factor 
affecting the $P_T$ slope is close to 1.6. As announced, the CEM yields start 
to  depart from the data when $P_T$ increases 
(see \cf{fig:single-Jpsi-pT-spetrum}), both at LO and NLO. This confirms
that the CEM can indeed be seen as an upper theory limit for $J/\psi$ production
processes dominated by gluon-fragmentation channels.

\section{$J/\psi+Z$ production at the LHC}

Cross section predictions for the associated production of a $J/\psi$ and a $Z$ boson 
at the LHC were provided up to NLO accuracy 
in \cite{Mao:2011kf,Gong:2012ah}. However, ATLAS found out~\cite{Aad:2014kba} 
larger SPS yields  than expected if the DPS rates were assumed
to be compatible with jet-related observables, \ie\ with  $\sigma_{\rm eff}$ 
on the order of 15 mb.

\begin{figure}[hbt!]
\begin{center}
\subfloat{\includegraphics[width=0.45\columnwidth]{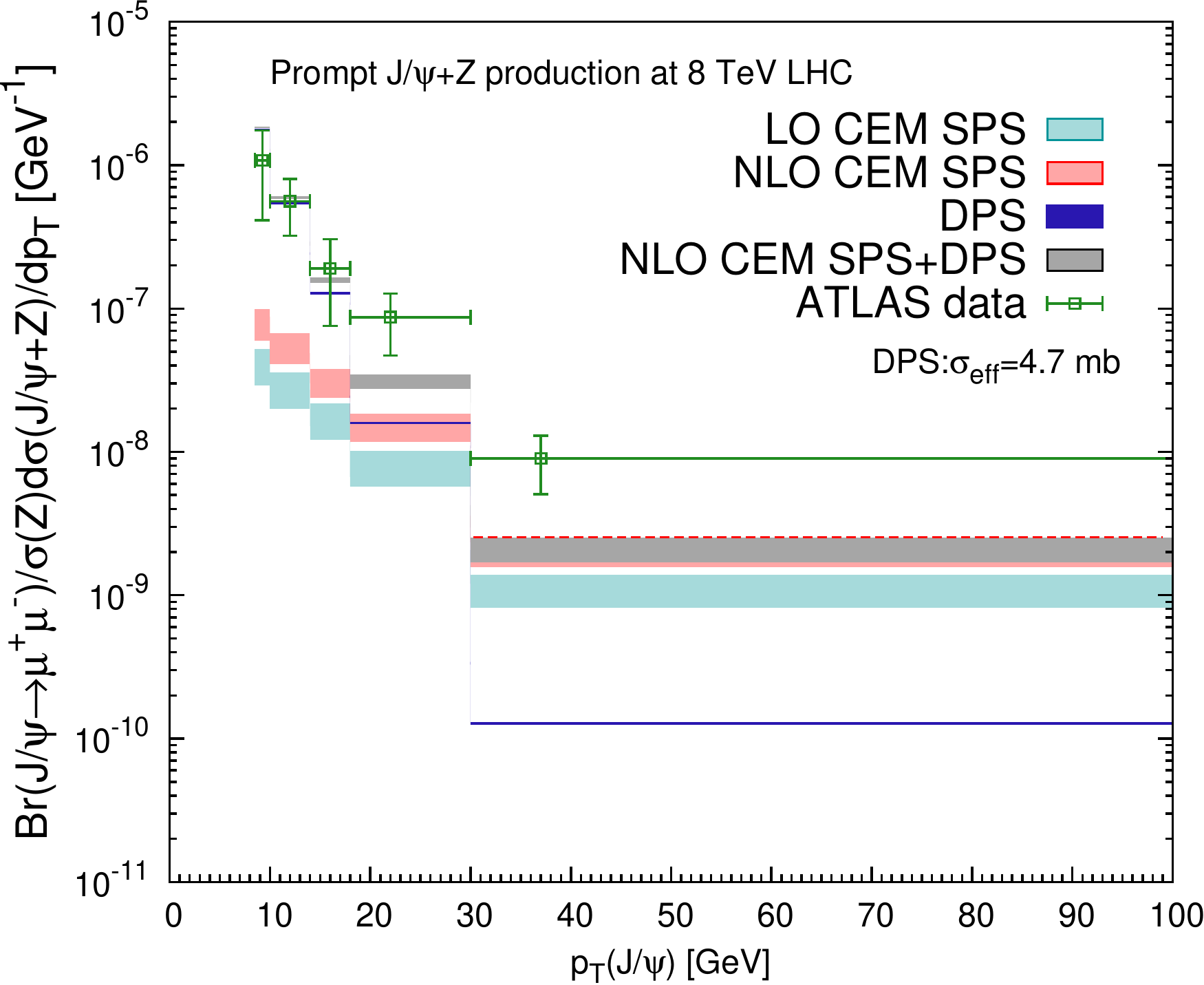}\label{fig:psi-Z-pt-1}}
\subfloat{\includegraphics[width=0.45\columnwidth]{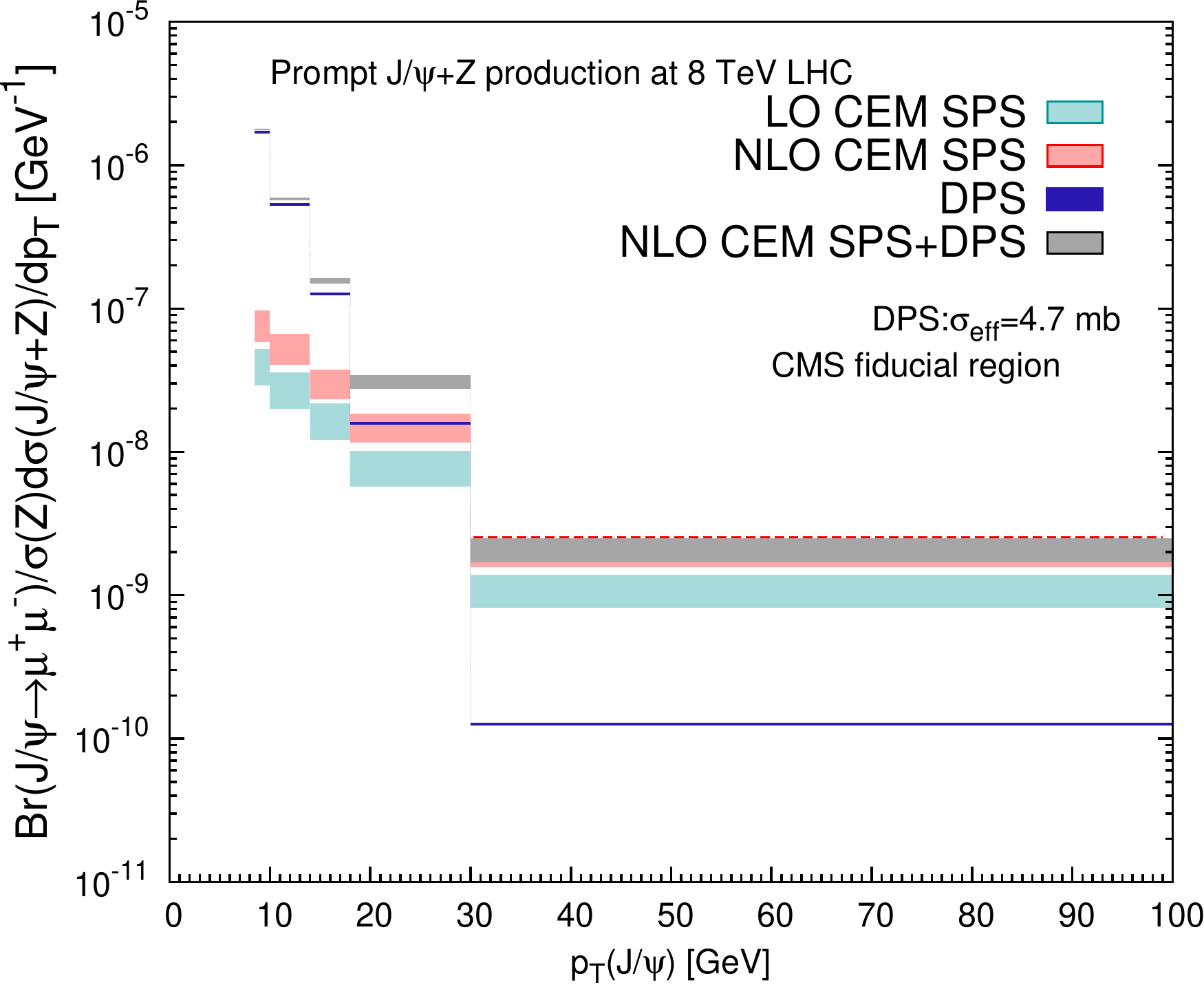}\label{fig:psi-Z-pt-1}}
\caption{(a) Comparison between the ATLAS $P_T^{J/\psi}$-differential cross section
 and our theoretical results for $J/\psi+Z$ at NLO CEM SPS + DPS. (b) Idem for the CMS acceptance (see~\cite{Lansberg:2016rcx} for details).
[Plots from~\cite{Lansberg:2016rcx}].
\vspace*{-.25cm}\label{fig:psi-Z-pt}}
\end{center}
\end{figure}

In~\cite{Lansberg:2016rcx}, we have shown that it would very unlikely that
the SPS contributions from any sensible approach would be compatible with the DPS-subtracted data 
using DPS contributions assuming $\sigma_{\rm eff}=15$ mb. Indeed, the CEM yield (computed up to NLO), which we consider to be 
an upper limit of the SPS contributions, does not agree with such an assumption on the DPS yield.
We further showed that the data could accomodate a $\sigma_{\rm eff}$ as low 
as 5 mb, which in turn gives a smaller DPS-subtracted yield closer to the SPS theoretical expectations. Yet, 
additional data are welcome to draw firmer conclusions. Comparisons with the ATLAS data are shown on \cf{fig:psi-Z-pt} for
the $P_T^{J/\psi}$-differential cross section (along with predictions for the CMS acceptance) and for the
(uncorrected) azimuthal distribution on \cf{fig:psi-Z-phi}.

\begin{figure}[hbt!]
\begin{center}
\subfloat{\includegraphics[width=0.45\columnwidth]{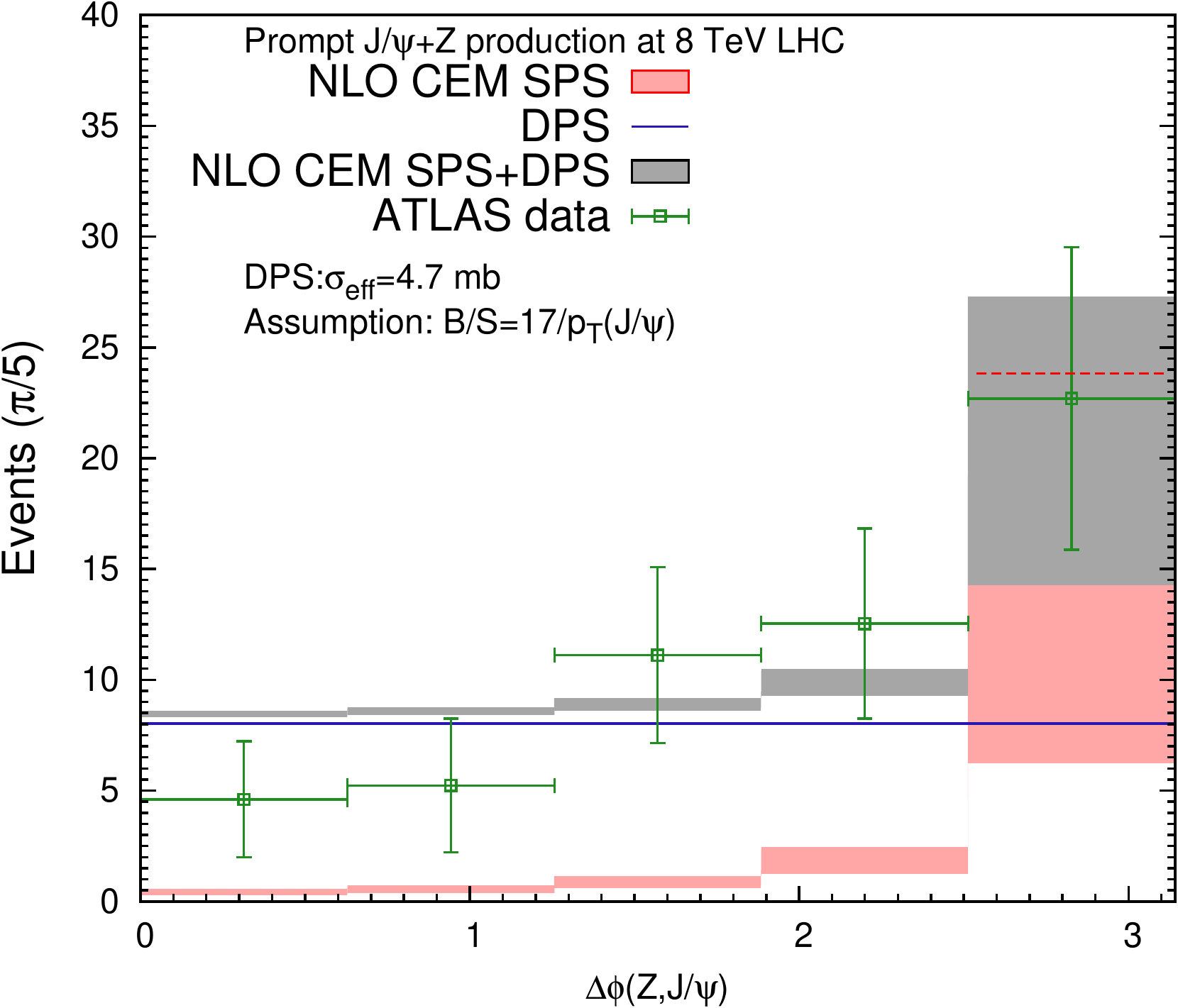}\label{fig:psi-Z-phi}}
\subfloat{\includegraphics[width=0.55\columnwidth]{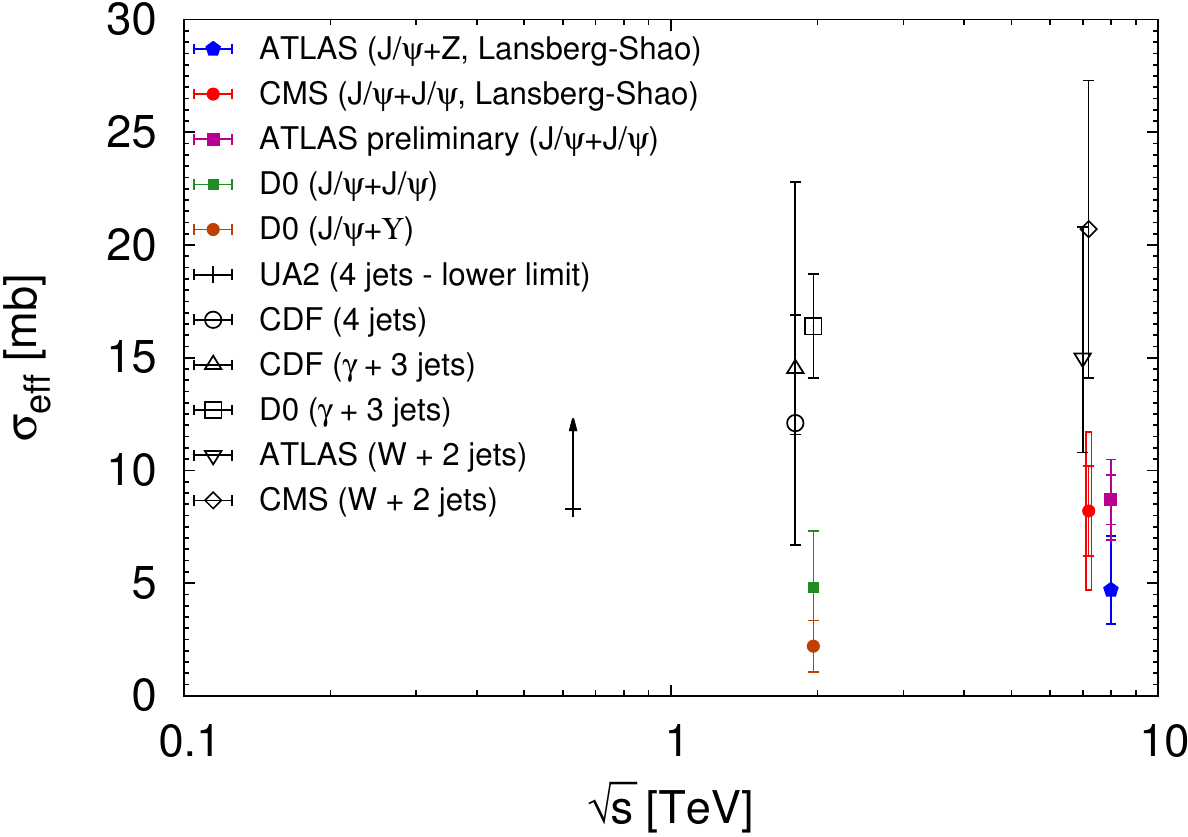}\label{fig:sigma_eff}}
\caption{(a) Comparison between the (uncorrected) ATLAS azimuthal event distribution
 and our theoretical results for $J/\psi+Z$ at NLO CEM SPS + DPS
effectively folded with an assumed ATLAS efficiency (see~\cite{Lansberg:2016rcx} for details). (b) Our ranges for $\sigma_{\rm eff}$ 
 extracted from the $J/\psi+Z$ data ($4.7^{+2.4}_{-1.5}$~mb)~\cite{Lansberg:2016rcx} 
and from di-$J/\psi$ data~\cite{Lansberg:2014swa} ($8.2\pm 2.0 \pm 2.9$~mb)
compared 
with 
other extractions~\protect\cite{Akesson:1986iv,Alitti:1991rd,Abe:1993rv,Abe:1997xk,Abazov:2009gc,Aad:2013bjm,Chatrchyan:2013xxa,Abazov:2014qba,ATLAS-CONF-2016-047}.[Plots from~\cite{Lansberg:2016rcx}.]}
\end{center}
\end{figure}

\section{Conclusion}

In the recent semesters, a significant number of experimental studies of associated-production of quarkonia have been lately carried out. 
We have reviewed one of them: the production of a $Z$ boson along with a prompt $J/\psi$. We have found
that the DPS contributions are indispensable to describe the data, pointing at a somewhat 
small $\sigma_{\rm eff}$ (see~\cf{fig:sigma_eff}) as do most of the other quarkonium-associated-production observables.

\section*{Acknowledgements}
The work of J.P.L. is supported in part by the French CNRS via the LIA FCPPL (Quarkonium4AFTER) and the D\'efi
Inphyniti-Th\'eorie LHC France. H.S.S. is supported by the ERC grant 291377 {\it LHCtheory:
Theoretical predictions and analyses of LHC physics: advancing the precision frontier}.

\bibliography{LANSBERG_JeanPhilippe_CONF12}

\end{document}